\documentclass[doublecol,final]{epl2}
\usepackage{amssymb}
\usepackage{amsmath}
\usepackage{bbm}
\usepackage{graphicx}

\begin{document}
\author{Emanuel Gull \inst{1} \and Philipp Werner \inst{2} \and Xin Wang \inst{2} \and Matthias Troyer \inst{1} \and A. J. Millis \inst{2}}

\institute{                    
  \inst{1} Theoretische Physik, ETH Zurich, 8093 Zurich, Switzerland\\
  \inst{2} Columbia University, 538 West, 120th Street, New York, NY 10027, USA \\
}
\title{Local Order and the gapped phase of the Hubbard model: a plaquette dynamical mean field investigation}

\date{\today}

\hyphenation{}

\abstract{
The four-site  ``DCA'' method of including intersite correlations
in the dynamical mean field theory is used to investigate the
metal-insulator transition in the Hubbard model.  At half filling a gap-opening transition
is found to occur as the interaction strength is increased beyond a critical value. 
The  gapped behavior  found in the 4-site DCA approximation 
is shown to be associated with the onset of
strong antiferromagnetic and singlet correlations and the transition
is found to be potential energy driven. It is thus more accurately described as a Slater phenomenon
(induced by strong short ranged order)
than as a Mott phenomenon. Doping the gapped phase leads
to a non-Fermi-liquid state with a Fermi surface only in the nodal 
regions and a pseudogap in the antinodal regions at lower dopings $x \lesssim 0.15$ and to a
Fermi liquid phase at higher dopings.
}

\pacs{71.30.+h}{Metal-insulator transitions and other electronic transitions}
\pacs{71.27.+a}{Strongly correlated electron systems; heavy fermions }
\pacs{71.10.Fd}{Lattice fermion models (Hubbard model, etc.) }
%\pacs{02.70.Ss}{Quantum Monte Carlo methods }
%\pacs{71.30.+h,71.27.+a,71.10.Fd}

\maketitle

Understanding the ``Mott" or correlation-driven metal insulator transition is one
of the fundamental questions in electronic condensed matter physics \cite{Mott49,Imada98}. Interest
increased following   P. W. Anderson's proposal that the copper oxide based high temperature
superconductors are doped ``Mott insulators" \cite{Anderson87}.  \footnote{
It is sometimes useful
to distinguish ``Mott" materials in which the important interaction scale is set directly by an interorbital Coulomb
repulsion  from ``charge transfer" materials  in which the interaction scale is set indirectly
via the energy required to promote a particle to another set of orbitals \cite{Zaanen85}. For present
purposes the difference is not important; the term Mott insulator will be used for both cases.
}

Clear theoretical pictures exist  in the limits of  strong and  weak coupling.  In strong coupling, insulating behavior results from the  ``jamming" effect  \cite{Mott49} in which  the presence of one electron in a unit cell blocks a second electron from entering; we term this the Mott mechanism.    At weak coupling, insulating behavior arises because long-ranged \cite{Slater51} or local \cite{Lee73,Kyung06} order opens a gap; we term this the Slater mechanism. (Anderson \cite{Anderson97} has argued that in 2d the strong coupling regime provides the appropriate description of the low energy behavior for all interaction strengths, but this view is controversial and does not address the question of interest here, namely  the physical origin of the novel low energy physics.) Many materials \cite{Imada98} including, perhaps,  high temperature superconductors \cite{Comanac08}
seem  to be in the intermediate coupling regime in which theoretical understanding is incomplete. 

The development of dynamical mean field theory, first in its single-site form \cite{Georges96} and subsequently in
its cluster extensions \cite{Hettler98,Kotliar01,Maier04,Okamoto03,Fuhrmann07} offers a mathematically well-defined approach to study metal-insulator transitions.
The method, while approximate, 
is non-perturbative and provides access to the intermediate coupling regime. In this paper 
we exploit new algorithmic developments \cite{Werner06,CTAUX} to obtain detailed solutions
to the dynamical mean field equations for the one orbital
Hubbard model in two spatial dimensions. This, the  paradigmatic model for the correlation-driven
metal-insulator transition, is  defined by the Hamiltonian
\begin{equation}
H=\sum_{p,\alpha} \varepsilon_pc^\dagger_{p,\alpha}c_{p,\alpha}+U\sum_in_{i,\uparrow}n_{i,\downarrow}
\label{H}
\end{equation}
with local repulsion $U>0$. We use the electron dispersion $\varepsilon_p=-2t(\cos p_x+\cos p_y)$.  
The dynamical mean field approximation to this model has been previously considered
\cite{Georges96,Maier04,Civelli05,Macridin06,Zhang07,Chakraborty07,Park08};
we comment on the differences to our findings below and in the conclusions.

The dynamical mean field method approximates the   electron self energy $\Sigma(p,\omega)$ by
\begin{equation}
\Sigma(p,\omega)=\sum_{a=1...N}\phi_a(p)\Sigma_a(\omega).
\label{dmftansatz}
\end{equation}
The  $N$ functions $\Sigma_a(\omega)$ are the self energies of 
an $N$-site quantum impurity model whose form is specified by a self-consistency condition. 
Different implementations of dynamical mean field theory 
correspond to different choices of basis functions $\phi_a$ and different self-consistency conditions
\cite{Okamoto03,Fuhrmann07}. In this paper we will use
primarily the ``DCA" ansatz \cite{Hettler98} although we
have also used the CDMFT method \cite{Kotliar01,Park08}  to verify our results and make comparison to other work.
In the DCA method one tiles the Brillouin zone into $N$ 
regions, and chooses $\phi_a(p)=1$ if $p$ is contained
in region $a$ and $\phi_a(p)=0$ otherwise. The   
``cluster momentum'' sectors  $a$ correspond roughly to averages of the corresponding
lattice quantities over the momentum regions in which $\phi_a(p)\neq 0$.
%In the case $N=1$ (corresponding to the original single-site dynamical mean field theory \cite{Georges96}) the full self energy is replaced by one function representing in effect the average of the self energy over the entire Brillouin zone. 

We present results for $N=1$ (single-site DMFT)  and $N=4$. Because we are interested
in the effects of short ranged order, the restriction to small clusters is not a crucial limitation:  while the precise 
parameter values at which the transition to insulating behavior occurs depend on cluster size, the 
basic relation we establish between correlations and the insulating behavior does not, and the 4-site cluster is computationally manageable 
so a wide range of information can be extracted. 

In the $N=4$ case
the impurity model is  a 4-site cluster  in which the cluster  electron creation operators $d^\dagger$
may be labeled either by a site
index $j=1,2,3,4$ or  by a cluster momentum variable $A=S,P_x,P_y,D$ with
$S$ representing an average over the range $(-\pi/2<p_x<\pi/
2;-\pi/2<p_y<\pi/2)$, $P_x$ over the range $(\pi/2<p_x<3\pi/2;-\pi/2<p_y<\pi/2)$, and $D$
over the range $(\pi/2<p_x<3\pi/2;\pi/2<p_y<3\pi/2)$.
The cluster states  are coupled to a bath of noninteracting electrons labeled
by the same quantum numbers. The Hamiltonian is
\begin{eqnarray}
H_\text{QI}&=&H_\text{cl}+\sum_{A,\sigma,\alpha}\left(V^\alpha_{A}d^\dagger_{A,\sigma}c^\alpha_{A,\sigma}+H.c.\right)+H_\text{bath}
\label{HQI},\\
H_\text{cl}&=&\sum_{A,\sigma}\varepsilon_A\left(d^\dagger_{A,\sigma}d_{A,\sigma}+H.c.\right)+U\sum_jn_{j,\uparrow}n_{j\downarrow}.
\label{Hcl}
\end{eqnarray}
We solve the impurity models on the imaginary frequency axis using two new continuous-time
methods \cite{Werner06,CTAUX}. Because we are studying a two dimensional model at temperature 
$T>0$ we restrict attention to phases without long ranged order.
The $\varepsilon_A$, $V^\alpha_A$ and $H_\text{bath}$ are determined by a self consistency condition \cite{Georges96,Fuhrmann07}.

\begin{figure}[tb]
\includegraphics[width=0.8\columnwidth]{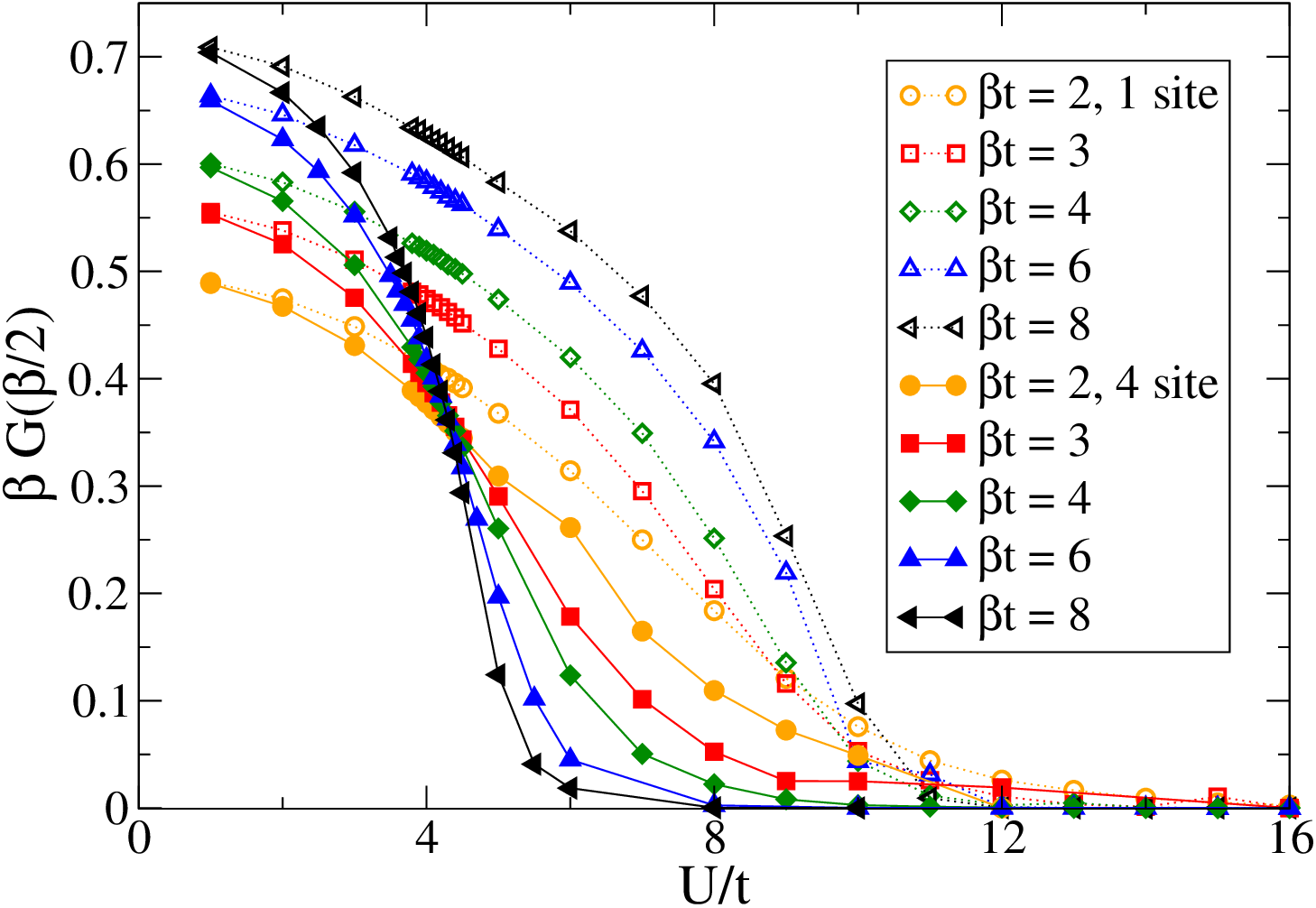}
\caption{On-site Green function at time $\tau=\beta/2$ computed using single-site and 4-site
DCA methods. All computations are performed in the paramagnetic phase at half filling.  }
\label{Gbeta2}
\end{figure}

The $N=1$ case has been extensively studied \cite{Georges96}.
At $N=1$, intersite correlations are entirely neglected;
the only physics is the strong correlation ``local blocking" effect envisaged by Mott.  
If attention is restricted to the paramagnetic phase,
to temperature $T=0$, and density $n=1$ per site  one finds that
the ground state is metallic for $U<U_{c2} \approx 12t$ \cite{Comanac08} and insulating for $U>U_{c2}$.  The
insulating phase is paramagnetic and characterized by an entropy of $\ln 2$ per site corresponding to the
spin degeneracy of the localized electrons. For
$U_{c1} \approx 9t<U<U_{c2}$ the insulating
phase, although not the ground state, is metastable and the extensive entropy of the insulating state leads
to a  transition to the insulating state as the temperature is raised \cite{Georges96}. 

The antiferromagnetic solution of the single-site DMFT equations 
has also been extensively studied. The model considered here has a nested Fermi surface at carrier
concentration $n=1$, so at $n=1$ the ground state is an insulating antiferromagnet at all interaction strengths $U$.
The N\'{e}el temperature peaks at $U\approx 0.8U_{c2}$ \cite{Comanac08}. This correlation strength also marks a change in
the character of the transition: for $U\lesssim 0.8 U_{c2}$  the expectation value
of the interaction term $Un_\uparrow n_\downarrow$ decreases as the magnetic order increases. 
The transition is thus potential energy driven and is identified with Slater physics. 
However for $U\gtrsim 0.8 U_{c2}$ the expectation value of the interaction term increases
as the system enters the antiferromagnetic phase; the transition in this case is thus kinetic energy driven
and is identified with Mott physics.

%\begin{figure}[t]\includegraphics[width=0.8\columnwidth]{phasediagram.eps}\caption{Metal-insulator phase diagram calculated at density $n_\alpha=0.5$ in plane of interaction strength $U$ and temperature $T$ using single site and 4-site DCA approximations. Also shown is the antiferromagnetic phase boundary and the single-site DMFT phase diagram. EMANUEL CAN YOU MAKE THIS FIGURE--SHOWING AN ESSENTIALLY VERTICAL PHASE BOUNDARY BETWEEN METAL AND INSULATING STATES, ALONG WITH B OTH THE PARAMAGENTIC AND ANTIFERROMAGNETIC PHASE BOUNDARIES. THESE ARE IN ARMIN'S THESIS. IF YOU DONT HAVE THIS I CAN SEND IT TO YOU.  }\label{phasediagram}\end{figure}

We now present results for the $N=4$ model in comparison to those obtained in the single-site approximation.
Figure~\ref{Gbeta2} presents the imaginary time Green function $G(R,\tau)$ at the particular values $R=0$
and $\tau=1/2T\equiv \beta/2$, computed at density $n=1$ per site for different temperatures $T$
and interactions $U$ using $1$ and $4$ site DCA. $G(0,\beta/2)$ is directly 
measured in our simulations and is related to the 
on-site electron spectral function $A_0(\omega)$ by
\begin{equation}
G(0,1/(2T))=\int \frac{d\omega}{\pi} \frac{A_0(\omega)}{2\cosh\frac{\omega}{2T}}\approx TA_0(\omega=0).
\label{Gbetahalf}
\end{equation}
The last approximate equality applies for sufficiently small $T$ and shows that the behavior of $G(0,\beta/2)$
provides information on the existence of a gap in the system. For $N=1$ and $U\lesssim 10t$ $G(0,\beta/2)$
increases as $T$ decreases, indicating the development of a coherent Fermi liquid state. 
In the 4-site DCA results a transition is evident as $U$ is increased through  $U^* \approx 4.2t$: for $U<U^*$  
$A(0)$ increases slowly as $T$ is decreased, 
as  in the single site model,
but for  $U>U^*$, $A(0)$ decreases, signaling the opening of a gap. 
%The very rapid change across $U=U^*$ suggests that the transition might be first order, 
%and the critical $U$ is seen to be essentially independent of temperature.
%(Park {\it et al.} have carefully studied the $T$-dependence of the phase boundary
%using the CDMFT method \cite{Park08}).
The very rapid change across $U=U^*$ is consistent with a first order transition,
as found in the careful CDMFT analysis of Park {\it et al.} \cite{Park08}.
The critical $U$ is seen to be essentially independent of temperature indicating that
the entropies of the metallic and non-metallic states are very similar.
The end-point of the first order
transition is at about $T=0.25t$ which is approximately the N\'{e}el temperature of the single-site 
method, at $U=4t$ \cite{Comanacthesis}.

\begin{figure}[tb]
\includegraphics[width=0.8\columnwidth]{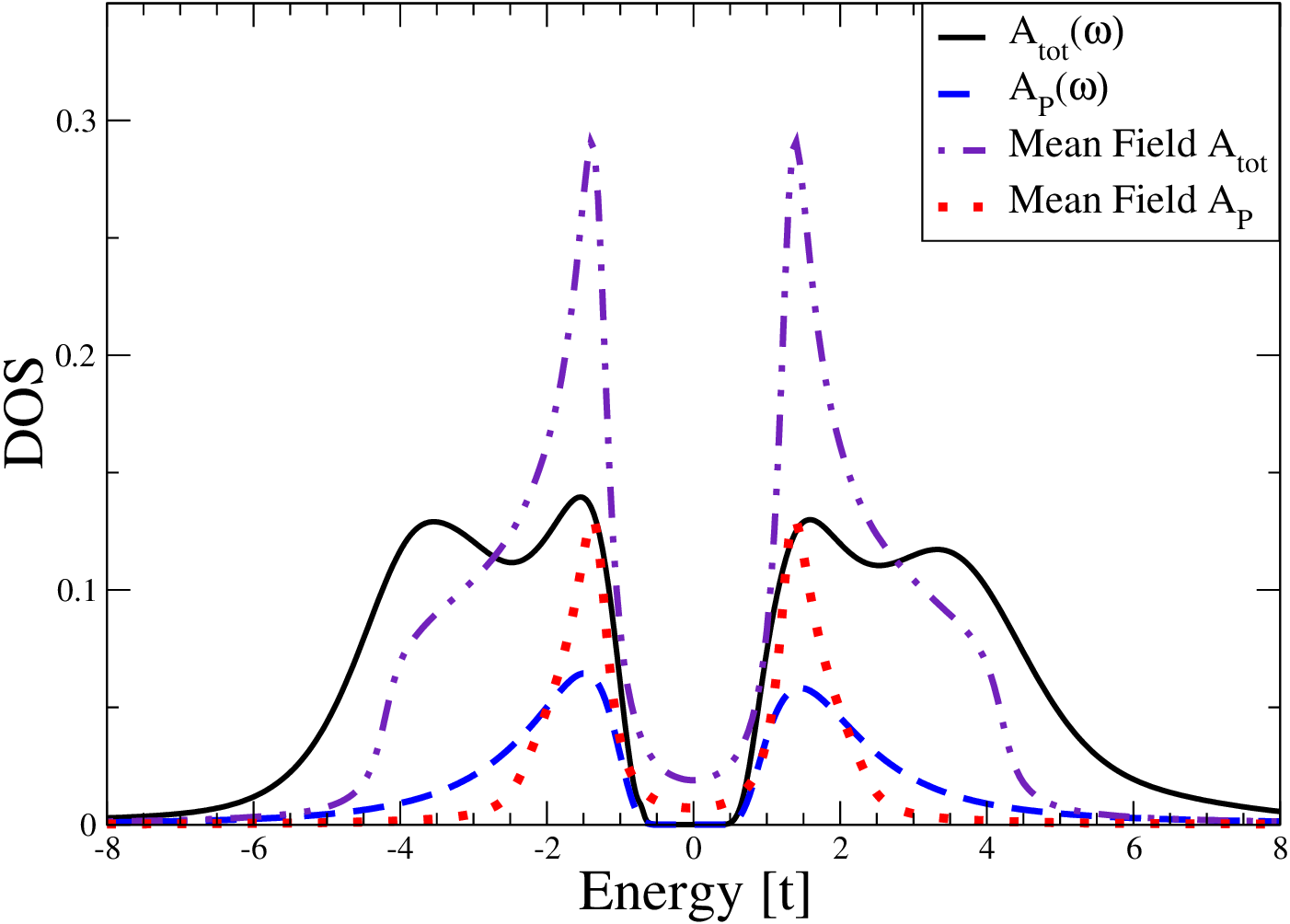}
\caption{Solid line: on-site spectral function computed by maximum entropy analytical
continuation of %CT-AUX 
QMC data for $U=6t$ and doping $x=0$. Dashed line: spectral function
in the $P=(0,\pi) , (\pi,0)$-momentum sector. Dotted and dash-dotted lines: $P=(0,\pi) , (\pi,0)$ and 
%local spectral functions obtained by a mean field calculation 
local spectral functions obtained by performing the DCA momentum averages of the 
standard SDW mean field expressions for the Green function,
with gap $\Delta=1.3t$. }
\label{spectral}
\end{figure}

Figure~\ref{spectral} shows as the solid line the local electron spectral function computed by 
maximum entropy analytical continuation of our QMC data for $U=6t$ and $n=1$.  
Analytical continuation is well known to be an ill-posed problem, with very small differences in
imaginary time data leading in some cases to very large differences in the inferred real axis quantities. 
A measure of the uncertainties in the present calculation comes from the
difference between the spectra in the  positive energy and negative energy regions, which
should be equal by particle-hole symmetry. We further note that the gap is consistent with the
behavior shown in Fig.~\ref{Gbeta2}.   The local spectral function exhibits a characteristic two-peak structure
found also in CDMFT calculations \cite{Park08}. 
The dotted line gives the spectral function for the $P_x$-sector, corresponding to an average of
the physical spectral function over the region $(\pi/2<p_x<3\pi/2),(-\pi/2<p_y<\pi/2)$; this is seen
to be the origin of the gap-edge structure. 

\begin{figure}[bt]
\includegraphics[width=0.8\columnwidth]{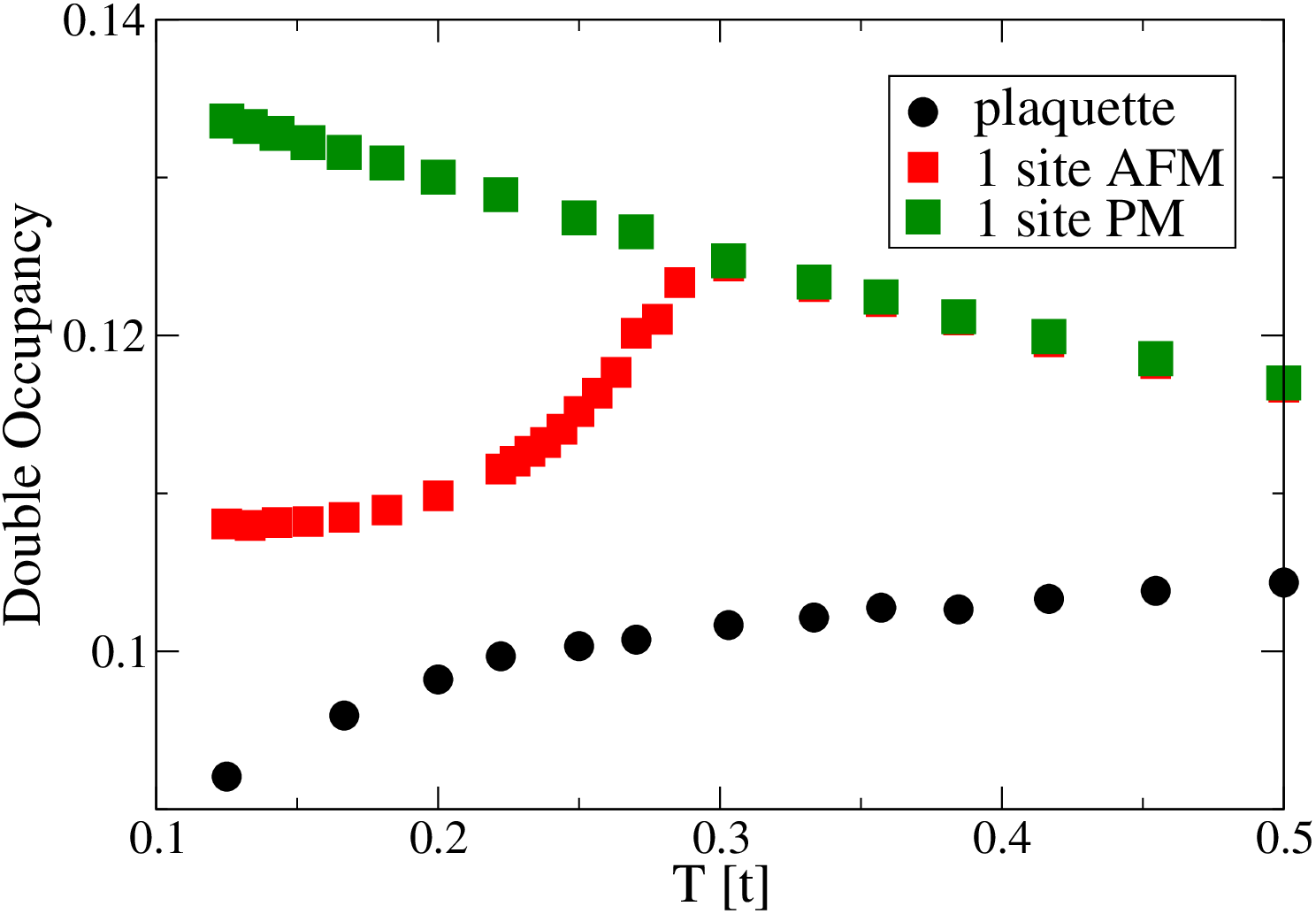}
\includegraphics[width=0.8\columnwidth]{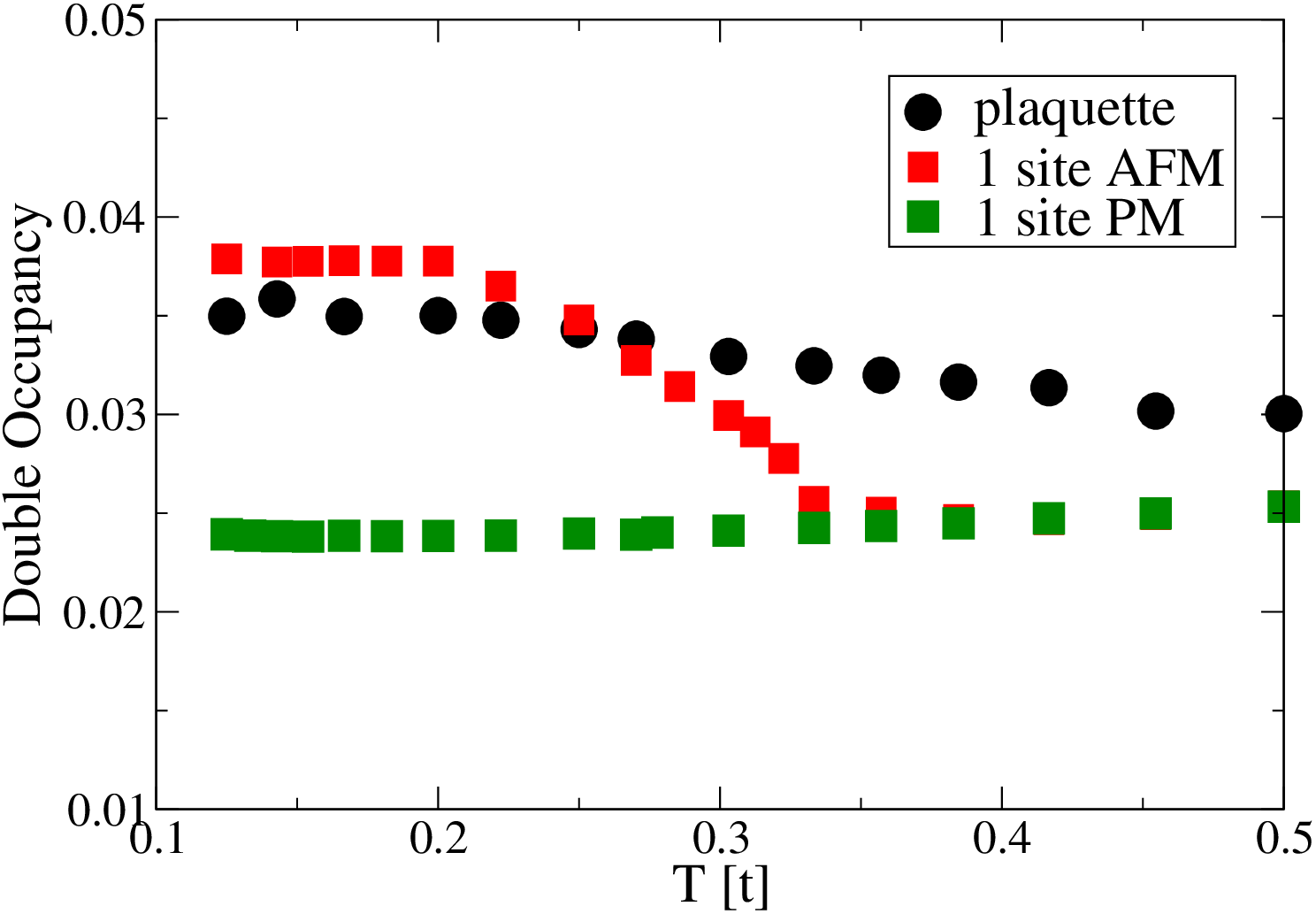}
\caption{Temperature dependence of 
double occupancy $\left<n_\uparrow n_\downarrow\right>$ computed using the 1-site
and 4-site DCA methods as a function of temperature for the half filled
Hubbard model at $U=5t$ (upper panel) and $U=10t$ (lower panel). The 1-site
calculations are done for both paramagnetic and antiferromagnetic phases whereas
the 4-site calculation is done for the paramagnetic phase only. }
\label{doubleoccupancy}
\end{figure}

We present in Fig.~\ref{doubleoccupancy}  the  temperature dependence of the 
double-occupancy $D=\left<n_\uparrow n_\downarrow\right>$ computed using 
the 1-site and  4-site DCA
for a relatively weak and a relatively strong correlation strength. 
In the single-site approximation antiferromagnetic correlations are absent in the
paramagnetic phase and become manifest  below the N\'{e}el temperature; the difference between
paramagnetic and antiferromagnetic phases therefore gives insight into the physics associated
with the antiferromagnetic correlations.   For the weaker interaction strength $U=5t$, the 
development of Fermi liquid coherence
as $T$ is decreased in the paramagnetic phase means that the wave function adjusts
to optimize the kinetic energy, thereby pushing the interaction term
farther from its extremum and  increasing $D$.  At this $U$ the 
magnetic transition is signaled by a rapid {\em decrease}
in $D$ , indicating that the opening of the gap 
enables a reduction of interaction energy, as expected if Slater physics dominates.
For  the larger $U=10t$ in the single site approximation we see that $D$ is temperature-independent
in the paramagnetic phase  because for this $U$ and temperature the model is in the Mott insulating state
(a first order transition to a metallic state would occur at a lower $T$). The antiferromagnetic transition
is signaled by an increase in $D$ because in the Mott state the transition to antiferromagnetism is kinetic energy driven.
%We present in Fig.~\ref{doubleoccupancy}  the  temperature dependence of the 
%double-occupancy $D=\left<n_\uparrow n_\downarrow\right>$ computed using 
%%single-site and 
%the 1-site and  4-site DCA
%for a relatively weak and a relatively strong correlation strength. 
%%inverse temperatures $\beta$.  
%In the single-site approximation to the paramagnetic phase
%and for the weaker interaction strength $U=5t$, the development of Fermi liquid coherence
%as $T$ is decreased means that the wave function adjusts
%to optimize the kinetic energy, thereby pushing the interaction term
%farther from its extremum and  increasing $D$.  At this $U$ the 
%magnetic transition is signaled by a rapid {\em decrease}
%in $D$ , indicating that the opening of the gap 
%enables a reduction of interaction energy, as expected if Slater physics dominates.
%For  the larger $U=10t$ in the single site approximation we see that $D$ is temperature-independent
%in the paramagnetic phase  because for this $U$ and temperature the model is in the Mott insulating state
%(a first order transition to a metallic state would occur at a lower $T$). The antiferromagnetic transition
%is signalled by an increase in $D$ because it is kinetic energy driven.

Turning now to the 4-site calculation we see at 
$U=5t$ a {\em decrease} in $D$ sets in below about $T^*=0.23t \approx 0.8 T_N^{\text{1-site}}$. $T^*$ is also the temperature below which $G(0,\beta/2)$ begins to drop sharply. This indicates that 
the opening of the gap  is related to 
a reduction of interaction energy, implying a ``Slater'' rather than a ``Mott'' origin
for the phenomenon.
For $U=10t$ we see a gradual  increase in $D$ as $T$ is decreased, reflecting the 
Mott physics effect of kinetic energy gain with increasing local 
antiferromagnetic correlations. 
% that for $U$ moderately greater than the critical value $U^*$  the double occupancy drops well  below the single-site value and continues to decrease as $T$ is decreased, signalling that the transition is of the Slater type.  On the other hand  for larger $U \gtrsim 10t$ the 4-site cluster results lie above the paramagnetic single-site results and increase as $T$ decreases, as expected if Mott physics is dominant. 

\begin{figure}[bt]
\includegraphics[width=0.8\columnwidth]{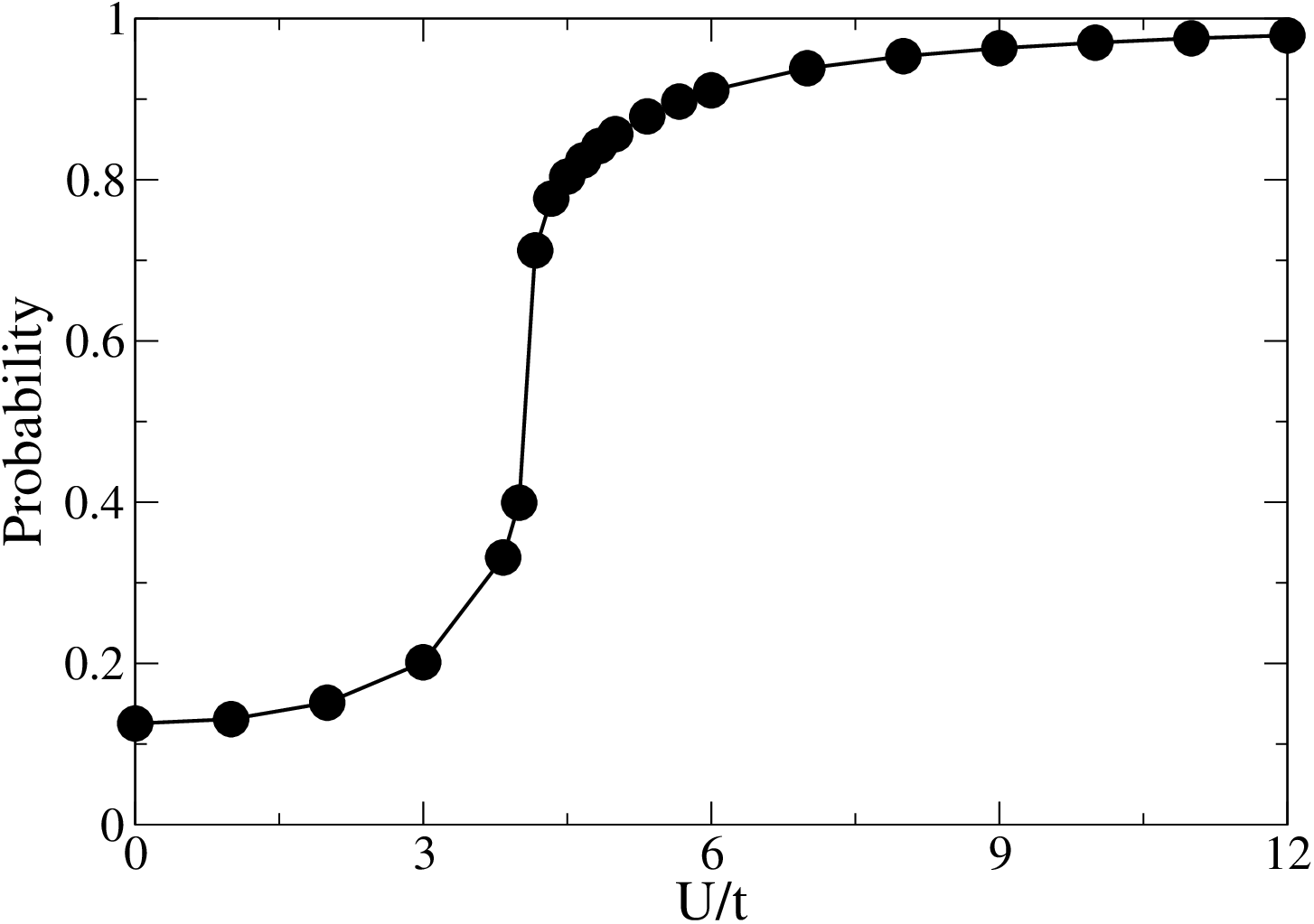}\\
%\vspace{6mm}
\includegraphics[width=0.8\columnwidth]{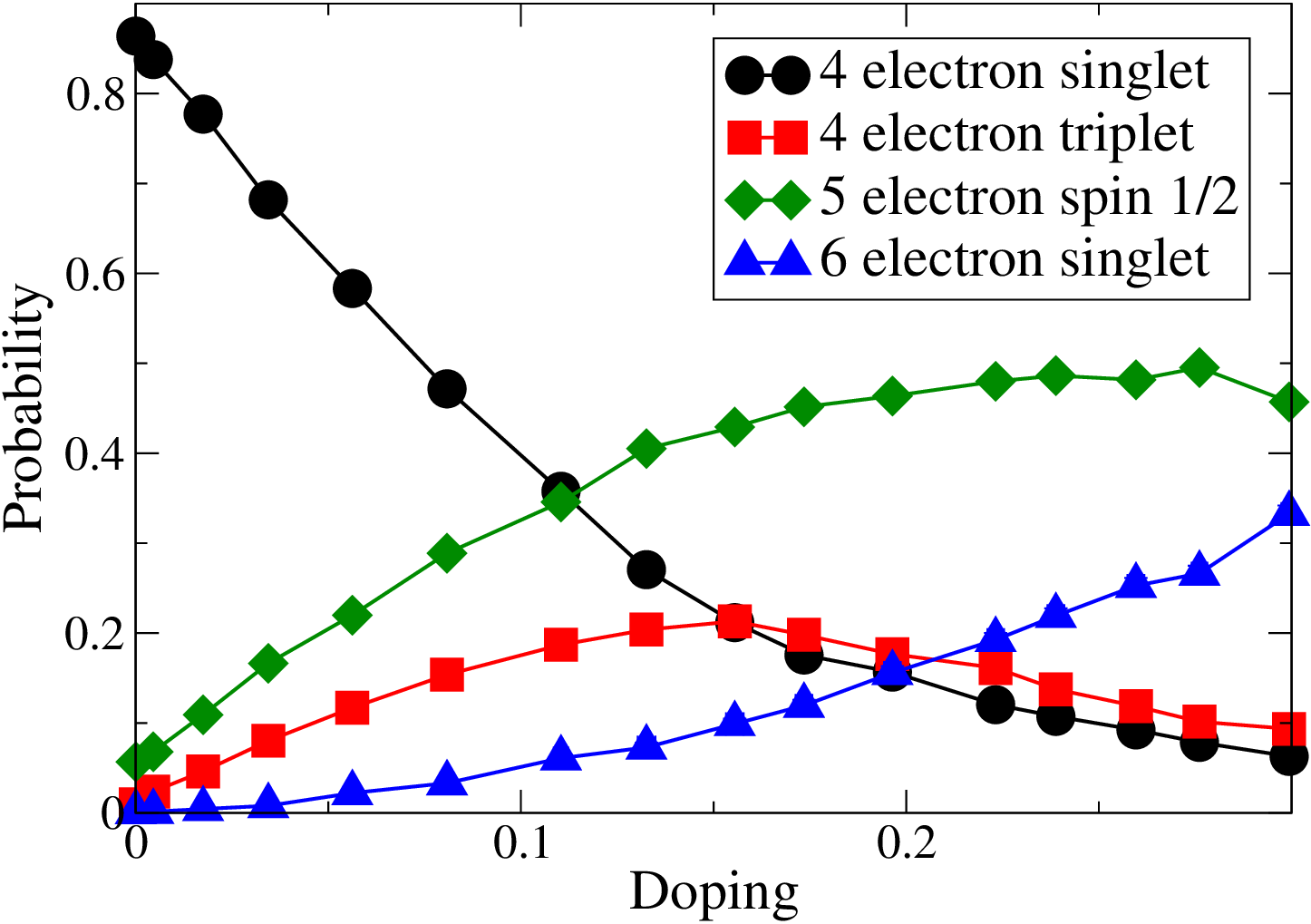}
\caption{
Probability that the local Hamiltonian is in a ``plaquette singlet state'' (a state with plaquette momentum $0$) at 
$n=1, T = t/30$, as a function of $U$. The sector statistics are measured in the hybridization algorithm.
%Upper panel: probability that the ``plaquette singlet" state is represented
%in the thermal ensemble at $n=1$, $T=t/30$ as a function of  $U$. 
Lower panel: evolution of the occupation probabilities with doping  at $U=5.2t$
and temperature $T=t/30$. }
\label{sectors}
\end{figure}

To further understand the physics of the transition we
%examine which eigenstates of $H_\text{cl}$ are
%represented with high probability in the actual state of the system.  
examine which eigenstates $|n_{cl}\rangle$ of $H_\text{cl}$ are
represented with high probability in the actual state of the system.  We define
$P_{n_{cl}}=\langle n_{cl}|{\hat \rho}_{cl}|n_{cl}\rangle$ with ${\hat \rho}_{cl}$ the cluster reduced density
matrix obtained by tracing the partition function over the bath states.
One particularly interesting state is the ``plaquette singlet" state which we denote
as $ |(12)(34)+(41)(23)\rangle$ with $(ab)$ representing a singlet bond between sites $a$ and $b$. 
The upper panel of Fig.~\ref{sectors}
shows the probability that this state is represented in the thermal ensemble corresponding to mean density 
$n=1$ for different interaction strengths $U$; the transition
at $U\approx 4.2t$ manifests itself as a dramatic
change (within our accuracy, the jump associated with a first order transition). 
We have performed CDMFT calculations to verify that that the same state and same physics
control the transition studied in Refs.~\cite{Zhang07,Park08}.

The plaquette singlet state has strong intersite correlations of both  $d$-wave  and  antiferromagnetic nature. 
It is natural to expect these correlations to open a gap in the electronic spectrum. To investigate this possibility
%we performed a mean field calculation of the lattice Green function using density $n=1$, a gap $\Delta=1.3t$
%and  antiferromagnetic and singlet pairing gaps
%and then used this in the DCA self consistency equation
%to obtain the impurity model spectral functions. 
we computed the DCA momentum averages of the lattice Green function using density $n=1$, 
and  antiferromagnetic and singlet pairing gaps of magnitude $\Delta=1.3t$
to obtain mean field estimates of the impurity model spectral functions. 
The dotted and dash-dotted lines
in Fig.~\ref{spectral} show the antiferromagnetic results. 
(Use of a $d$-wave pairing gap would yield very similar
results, except that instead of a clean gap at $0$ one finds a ``soft'' gap with a linearly vanishing density of states).
The evident similarity to the calculations
reinforces the argument that it is the local correlations
which are responsible for the gapped behavior. 
\begin{figure}[bt]
\includegraphics[width=0.8\columnwidth]{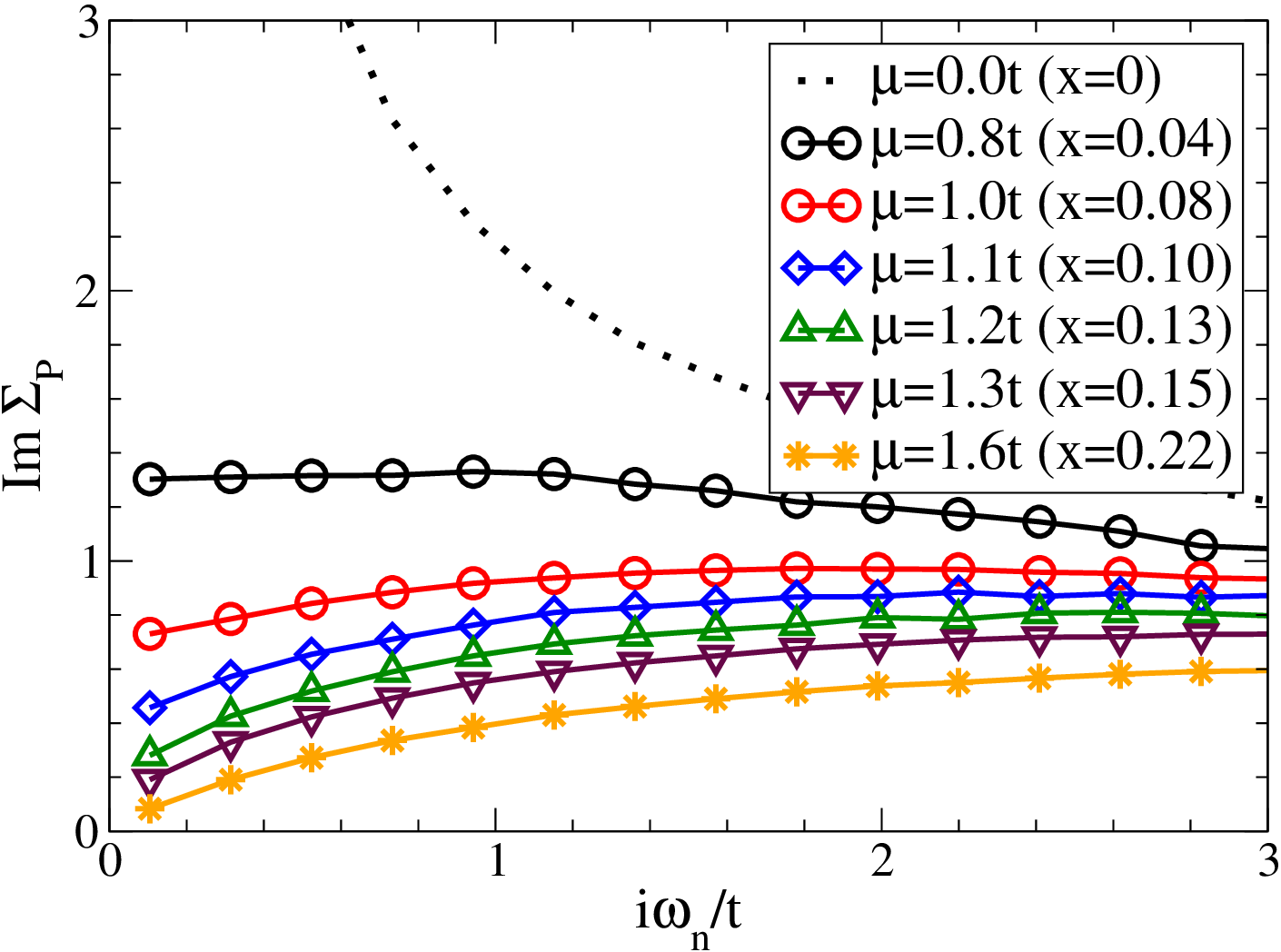}
\caption{Imaginary part of Matsubara-axis $P=(0,\pi) , (\pi,0)$-sector self energy measured
for $U=5.2t$ at temperature $T=t/30$ and chemical potential $\mu$ (doping $x$ per site) indicated. }
\label{selfenergy}
\end{figure}

We finally consider the effect of doping.
The model we study
is particle-hole symmetric. For definiteness we present results for electron doping.   
In a Fermi liquid, the imaginary part of the real-axis self energy is
$\text{Im} \Sigma(p,\omega\rightarrow 0) \propto \omega^2$. The spectral representation $\Sigma(i\omega_n)=\int \frac{dx}{\pi} \text{Im} \Sigma(p,x)/(i\omega_n-x)$  
then implies that at small $\omega_n$, $\text{Im} \Sigma(p,i\omega_n)\propto \omega_n$.
We find that in the $S=(0,0)$ and $D=(\pi,\pi)$ momentum sectors, this relation is obeyed at all dopings. The behavior in the $P=(0,\pi) , (\pi,0)$-sector
is different, as is shown in Fig.~\ref{selfenergy}. The dashed line shows the self energy for the half-filled model.
The $\omega_n^{-1}$ divergence, arising from the insulating gap, is evident. For large enough doping ($x\gtrsim 0.15$)
the expected Fermi liquid behavior is observed (and indeed for $x>0.2$ the self energy is essentially the
same in all sectors); however for smaller dopings, up to $x \approx 0.15$, $\text{Im} \Sigma_P$ 
does not extrapolate to $0$ as $\omega_n\rightarrow 0$,
indicating a non-Fermi-liquid behavior in this momentum sector. 

\begin{figure}[bt]
\includegraphics[width=0.8\columnwidth]{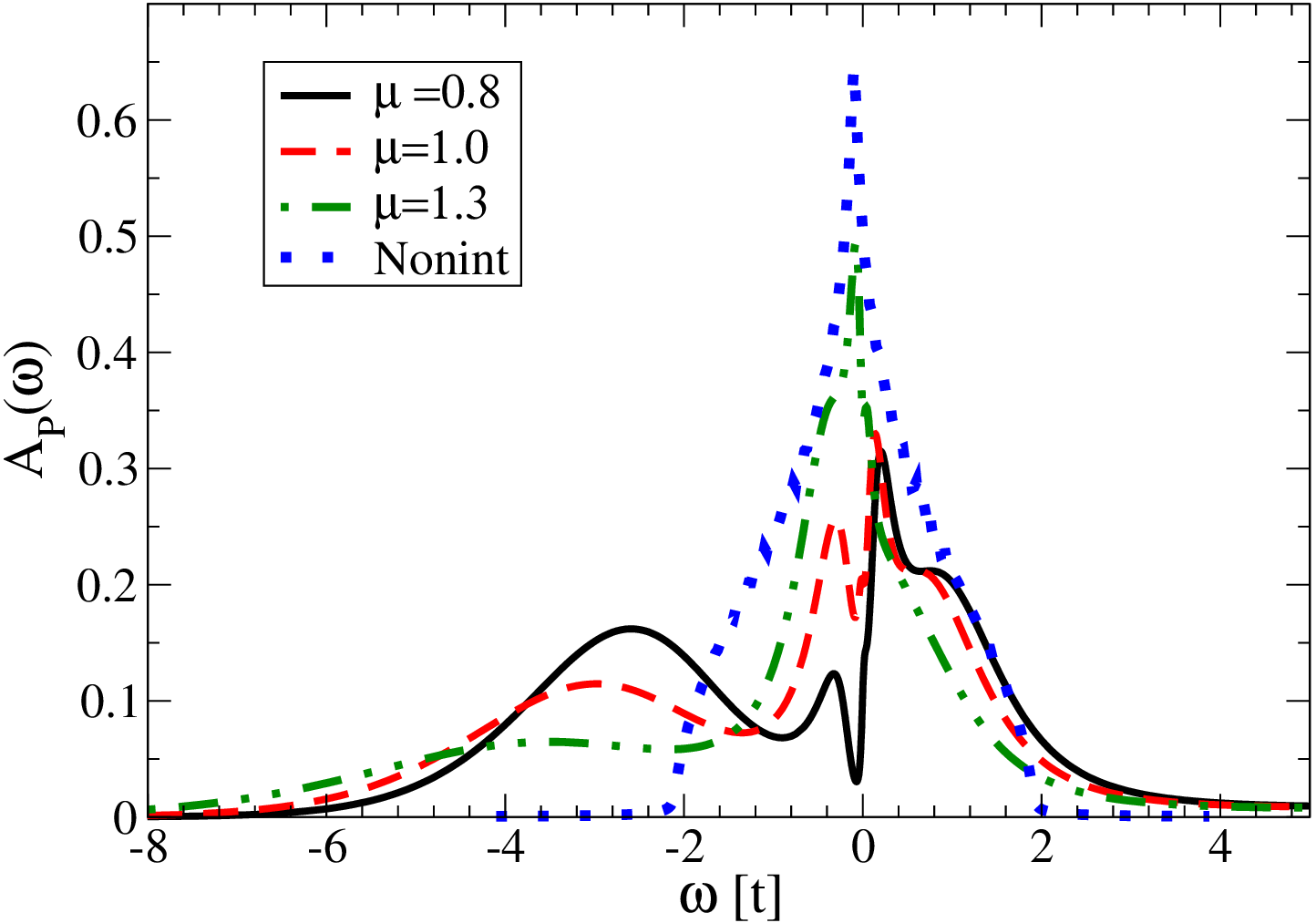}
\caption{Doping dependence of $P=(0,\pi) , (\pi,0)$-sector density of states
obtained by analytical continuation of quantum Monte Carlo data  at $U=5.2t$
and temperature $T=t/60$. }
\label{pdos}
\end{figure}

To explore further the non-Fermi-liquid behavior we present in Fig.~\ref{pdos}
the density of states in the $P=(0,\pi) , (\pi,0)$-sector, obtained by analytical continuation of our
quantum Monte Carlo data.  Comparison to Fig.~\ref{spectral} shows
that as the chemical potential is increased the Fermi level moves into the
upper of the two bands. In addition, for the lower dopings
a small `pseudogap' (suppression of density of states) appears
near the Fermi level while for $x=0.15$ the value of the spectral function
at the Fermi level approaches that of the noninteracting model, indicating
the restoration of Fermi liquid behavior. We have verified that these features
are robust, and in particular that the suppression of the density of states
near the Fermi level is required to obtain the measured values of $G(\tau \sim \beta/2)$.
Comparison of data obtained for inverse
temperature $\beta t=30$ and $\beta t=100$ (not shown) with the data obtained for $\beta t=60$ shown in Fig.~\ref{pdos}
is consistent with
the pseudogap being the asymptotic low-$T$ behavior, not an intermediate $T$ artifact.

Examination of the $D=(\pi,\pi)$-sector density of states and self energy shows
that for $x=0.04$ and $x=0.08$
there is no Fermi surface crossing in the $D=(\pi,\pi)$-sector, so within the 4-site DCA
approximation there is no Fermi surface at all. At these chemical potentials most doping is provided by incoherent, 
pseudogapped quasiparticles in the $P=(0,\pi) , (\pi,0)$-sector. 
As $x$ is increased beyond $\sim 0.1$ a Fermi crossing appears,
first in the $D$ sector and then for $x/\gtrsim 0.15$ in the $P$ sector, signaling
the restoration of Fermi liquid behavior.
The results may be interpreted as ``Fermi arcs" or as hole pockets
bounded by the edges of the $D=(\pi,\pi)$-sector: the momentum resolution
of the 4-site DCA is insufficient to distinguish the two. As the doping is further increased
the ``Fermi arc'' regions rapidly grow and the pseudogap fills in, leading to a
restoration of a conventional Fermi surface for $x>0.15$.  

The lower panel of Fig.~\ref{sectors} shows that this non-Fermi-liquid
behavior can be related to the prominence of  the plaquette singlet
and the plaquette triplet states.  The contribution of the plaquette triplet state peaks at $x\approx 0.15$, while
the contribution of the 6-electron singlet state remains small,
indicating a prominent role for antiferromagnetic (rather than $d$-wave singlet) correlations at this doping.
However, the increasing prominence of the 6-electron singlet state as doping is increased strongly suggests
that the larger doping Fermi-liquid-like state will be susceptible to a pairing instability. Similar results were found in CDMFT calculations 
by Kyung and collaborators \cite{Kyung06}, who attributed them to antiferromagnetic
correlations, by Zhang and  Imada \cite{Zhang07}   and by  Haule and Kotliar \cite{Haule07}.
 
In summary, we have shown that the insulating behavior (at doping $x=0$) and non-Fermi liquid behavior
(at doping $0<x<0.15$) found at relatively small $U$ in cluster dynamical mean field calculations
\cite{Civelli05,Macridin06,Zhang07,Park08,Kyung06,Moukouri01} may be understood
as a consequence of a potential-energy-driven transition to a state with definite, strong spatial correlations, mainly of the plaquette singlet type. 
Doping this state leads to a low energy pseudogap for momenta in the $P=(0,\pi) , (\pi,0)$ sector.   Superconducting
correlations (marked by the prominence of the 6 electron states) do not
become important until beyond the critical concentration at which Fermi liquid behavior
is restored. Our results are consistent with the
finding of Park {\it et. al.} \cite{Park08} 
that the $U$-driven transition is first order (although unlike those authors
we have not performed a detailed study of the coexistence region). 
We interpret
the transition as being driven by Slater (spatial ordering) physics, whereas 
Park {\it et. al.} interpret  their results as arising from a strong coupling, Mott phenomenon.
Moukouri and Jarrel \cite{Moukouri01} argue that Slater physics is
not important because in a 2d model with Heisenberg symmetry long range order does not set in until $T=0$; We believe, however,
that the results for double occupancy shown in Fig.~\ref{doubleoccupancy} and the dominance of particular states in the sector statistics plot 
Fig.~\ref{sectors} provide strong evidence that the physics is indeed dominated by local order, consistent with Slater-type physics.
The importance of spatial correlations for the spectral function and non-Fermi-liquid behavior was previously stressed
by Jarrell and co-workers \cite{Macridin06} and Zhang and Imada \cite{Zhang07}.
We also suggest that the short ranged order is responsible for
the features noted by Chakraborty and co-workers in the optical  conductivity and spectral 
function \cite{Chakraborty07}.  %The importance of spatial correlations was previously stressed by Jarrell and co-workers \cite{Maier04,Moukouri01,Macridin06}, Kyung et. al \cite{Kyung06}
%in a study of a similar model (but with second neighbor hoppings, a 16 site cluster
%and only density $n=1$ per site) \cite{Macridin06} and  and Zhang and Imada \cite{Zhang07}.  
Calculations in progress will extend the results presented here to larger clusters. 

{\it Acknowledgments} AJM and PW are  supported by  NSF-DMR-0705847 and EG and MT by the Swiss National Science Foundation.
All calculations have been performed on the Hreidar and Brutus clusters of ETHZ
using the ALPS \cite{ALPS} library.
\bibliographystyle{eplbib}

\end{document}